\pdfoutput=1

\documentclass[aps,pre,reprint,groupedaddress,floatfix]{revtex4-1}
\usepackage{graphicx}
\usepackage{epstopdf}

\usepackage{amssymb}
\usepackage{amsmath}
\usepackage{psfrag}
\usepackage{pstool}
\usepackage{layouts}

\newcommand{\Pe}{\textrm{Pe}\,}
\newcommand{\St}{\textrm{St}\,}

\begin{document}

\title{Thermal impact of magmatism in subduction zones}
\author{David W. \surname{Rees Jones}} 
\email[Corresponding author: ]{David.ReesJones@earth.ox.ac.uk}
\author{Richard F. \surname{Katz}}
\author{Meng \surname{Tian}}
\affiliation{Department of Earth Sciences, University of Oxford, South Parks Road, Oxford, OX1 3AN, UK.}
\author{John F. \surname{Rudge}}
\affiliation{Department of Earth Sciences, Bullard
Laboratories, University of Cambridge, Madingley Road, Cambridge CB3 0EZ, UK.}

\date{\today}

\begin{abstract}
Magmatism in subduction zones builds continental crust and causes most of Earth's subaerial volcanism. The production rate and composition of magmas are controlled by the thermal structure of subduction zones.  A range of geochemical and heat flow evidence has recently converged to indicate that subduction zones are hotter at lithospheric depths beneath the arc than predicted by canonical thermomechanical models, which neglect magmatism. We show that this discrepancy can be resolved by consideration of the heat transported by magma.  In our one- and two-dimensional numerical models and scaling analysis, magmatic transport of sensible and latent heat locally alters the thermal structure of canonical models by $\sim$300~K, increasing predicted surface heat flow and mid-lithospheric temperatures to observed values. We find the advection of sensible heat to be larger than the deposition of latent heat. Based on these results we conclude that thermal transport by magma migration affects the chemistry and the location of arc volcanoes. 
\end{abstract}

\maketitle

\section{Introduction}
Petrological estimates of sub-arc temperature conditions in both continental and oceanic subduction zones are systematically higher than predicted by thermal models, typically by 200--300~K, at depths less than $\sim$70~km  \cite{Kelemen2003,Perrin2016}. Similarly, measurements of geothermal heat flow in SW~Oregon and NE~Japan are higher than predicted by approximately 50--100~\mbox{mW/m$^2$} near the volcanic arc \cite{Kelemen2003,Furukawa1993}. Geophysical evidence from seismic and magnetotelluric imaging of high temperatures and/or magma at depth under volcanic arcs \cite{Zhao2007,Syracuse2008,Rychert2008,McGary2014} is consistent with the emerging consensus that the shallow arc temperatures in subduction zones are hotter than canonical models predict. 

In canonical models, the thermal structure of subduction zones is calculated as a balance between thermal diffusion and advection. Heat is advected by the creeping solid mantle flow within the wedge-shaped region between the subducting slab and overriding lithosphere \cite{McKenzie1969}. Previous modelling efforts to resolve the discrepancy with observations have involved varying the prescribed geometry of subduction, the coupling between mantle and slab, and the rheological model of the mantle \cite{Kelemen2003,Furukawa1993}. Inclusion of frictional heating along the slab top in the seismogenic zone increases heat flow in the fore-arc \cite{Gao2014}. None of these efforts have been successful in explaining both the amplitude of the thermal observations and their position relative to the volcanic arc. 

It is known that hydrous fluids are released from the subducting slab by de-volatilization reactions \cite{Schmidt2014} and percolate upward into the mantle wedge.  There they reduce the solidus temperature, promote melting, and hence become silicic as they ascend. During their ascent, the magmas traverse from cooler mantle adjacent to the slab, to hotter mantle at the core of the wedge, to cooler mantle at the base of the lithosphere. They advect heat between these regions and consume or supply latent heat with melting and freezing. Despite the copious production of magma in subduction zones, these thermal processes have been neglected from almost all previous models. One exception, a scaling argument comparing advective heat transport by magma flow to thermal diffusion, suggests that magma flow may be significant \cite{Peacock1990}. Similarly, hydrothermal circulation in the crust may play a role in cooling the slab in the fore-arc region \cite{Spinelli2016}. In this paper we assess the role of magmatic processes in altering the thermal structure of the wedge and lithosphere.  Our approach is based on theory for two-phase dynamics of the magma--mantle system \cite{McKenzie1984}. We quantify the magmatic  transport of sensible and latent heat, focusing on the physical mechanisms and their controls, rather than on any particular subduction zone.

\section{Methodology}
Magma migration in the mantle is a two-phase flow, governed by continuum equations of mass and momentum conservation for the solid (mantle) and melt (magma) \cite{McKenzie1984,Rudge2011}. The thermal and compositional structure is governed by equations of conservation of energy and chemical species. Our approach is to prescribe the magmatic flux and investigate how the thermal structure responds. This response is determined from energy conservation in the form of a heat equation:
\begin{equation}
\frac{\partial T}{\partial t} +\boldsymbol{v}_s  \cdot \nabla T + \boldsymbol{v}_D  \cdot \nabla T  = \kappa \nabla^2 T - \frac{L}{\rho c_p} \Gamma, \label{eq:heat}  
\end{equation}
$T$ denotes temperature, $t$ time, $\kappa$ thermal diffusivity, $\rho$ density, $c_p$ specific heat capacity, $L$ latent heat, and $\Gamma$ melting rate.  We neglect differences between the thermal properties of the phases because these do not affect the solution at leading order. The velocity variables involved are: solid mantle velocity $\boldsymbol{v}_s$,  liquid magma velocity $\boldsymbol{v}_l$,  the Darcy (or segregation) flux $\boldsymbol{v}_D\equiv\phi(\boldsymbol{v}_l-\boldsymbol{v}_s)$, where $\phi$ is the porosity. 

In the absence of magma, $\boldsymbol{v}_D=0$ and $\Gamma = 0$ and eqn.~\eqref{eq:heat} reduces to the heat equation used in canonical mantle convection calculations. In the presence of magma, two relevant terms are non-zero: first, an advective term associated with the segregation flux of magma $\boldsymbol{v}_D$; second, a latent heat sink associated with melting ($\Gamma>0$), which becomes a source in the case of freezing ($\Gamma<0$). The petrological model for $\Gamma$ is described in Sec.~S1, Supplementary Material, and was inspired by previous studies of mantle melting in the presence of water \cite{Hirschmann1999III,Katz2003,Keller2016}. 

By the considerations above and the results below, we emphasize that the latent heat of phase change is not the only thermal contribution from magmatism; there is also advective transport by the magma. In what follows, we consider the relative importance of these mechanisms.

\section{Results}
\subsection{One-dimensional model}
So-called `melting-column models' have been used to understand mid-ocean ridge magmatism, where the main cause of melting is decompression of the upwelling  mantle \cite{Ribe1985, Asimow1999, Hewitt2010}. Subduction zones are a considerably more complex environment, but we adapt ideas from melting-column models to investigate how magmatism modifies their thermal structure. The column model is fully derived and described in more detail in Sec.~S2, Supplementary Material. A one-dimensional, steady-state heat equation can be written
\begin{equation}  \label{eq:heat1D_dimensional}
\rho c_p \overline{W_0} \frac{d T}{dz} - \rho c_p \Psi^*  =  \frac{d }{dz}\left(\rho c_p \kappa \frac{d T}{dz}\right) -L \Gamma , 
\end{equation}
where $\Psi^*$ is the dimensional version of the source term, discussed below. 
We rescale lengths by the height of the column $H$, velocities by the diffusive scale \mbox{$\kappa/H$}, and $\Psi^*$ by \mbox{$\kappa/H^2$}. Then eqn.~\eqref{eq:heat1D_dimensional} becomes
\begin{equation}
  \Pe T' - \Psi = T'' -\Pe \St (T'+\Delta T_H), \label{eq:heat1D}  \\
\end{equation}
where $\Psi$ is the rescaled version of the source term, discussed below. 
$\Delta T_H$ is the adiabatic temperature drop between slab and surface; primes denote a derivative with respect to position (e.g., $T'$ is a rescaled vertical temperature gradient). Two dimensionless numbers control the behaviour of the system: a P\'{e}clet number $\Pe=H\overline{W_0} / \kappa$ is the scaled volume flux at the base of the column; a Stefan number $\St = (L/c_p) \partial F/\partial T$ is the scaled isobaric productivity that quantifies the ratio of latent to sensible heat ($F$ is the degree of melting). Hydrous flux melting has low isobaric productivity \cite{Hirschmann1999III} so the Stefan number is small. 

\begin{figure}[t]
\includegraphics[width=8.64cm]{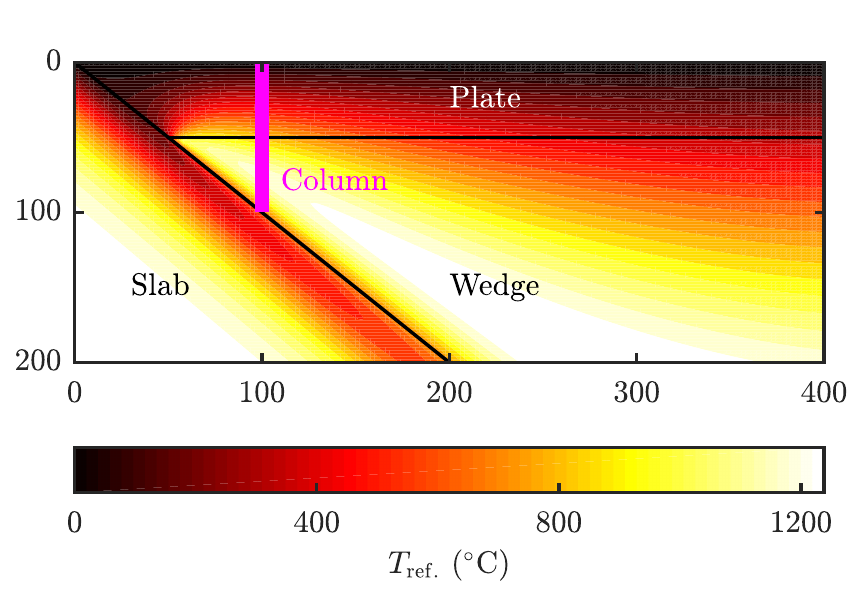}
\caption{Reference temperature field $T_\mathrm{ref.}$ from Ref.~\cite{vanKeken2008} using the parameter values listed therein. The dip angle, slab velocity and thickness of the overriding plate are prescribed. The solid velocity in the mantle wedge is calculated and coupled to the temperature through a temperature-weakening viscosity. A pink line indicates the position of an example column model. Axis label show distance from the trench in km. Only a subset of the model domain is shown; the full domain is 660~km wide and 600~km deep.} \label{fig:T_contour}
\end{figure}

The mantle flow in subduction zones is far from one-dimensional; a corner flow is driven by the motion of the subducting slab \cite{McKenzie1969}. A key step in representing corner flow in a column model is to introduce a spatially variable, volumetric heating term $\Psi$ that mimics the effects of large-scale mantle flow, which tends to supply heat into the column. We infer $\Psi$ from a single-phase, two-dimensional thermomechanical reference model that is shown in Figure~\ref{fig:T_contour}; the domain geometry and temperature-dependence of viscosity are as given in a study that outlined broadly representative models of subduction \cite{vanKeken2008}. From the reference model, we extract a vertical temperature profile at some position of interest $T_\mathrm{ref.}(z)$ and use it to calculate the source term $\Psi=-T''_\mathrm{ref.}$. The source term is constructed such that the solution of equation \eqref{eq:heat1D} in the absence of magma flow ($\Pe=0$) is $T=T_\textrm{ref.}$, i.e., the single-phase result. For $\Pe>0$, this approach is reasonable provided melt does not drastically change the large-scale mantle dynamics, a prospect we consider later. 

\begin{figure*}
\includegraphics[width=16.44cm]{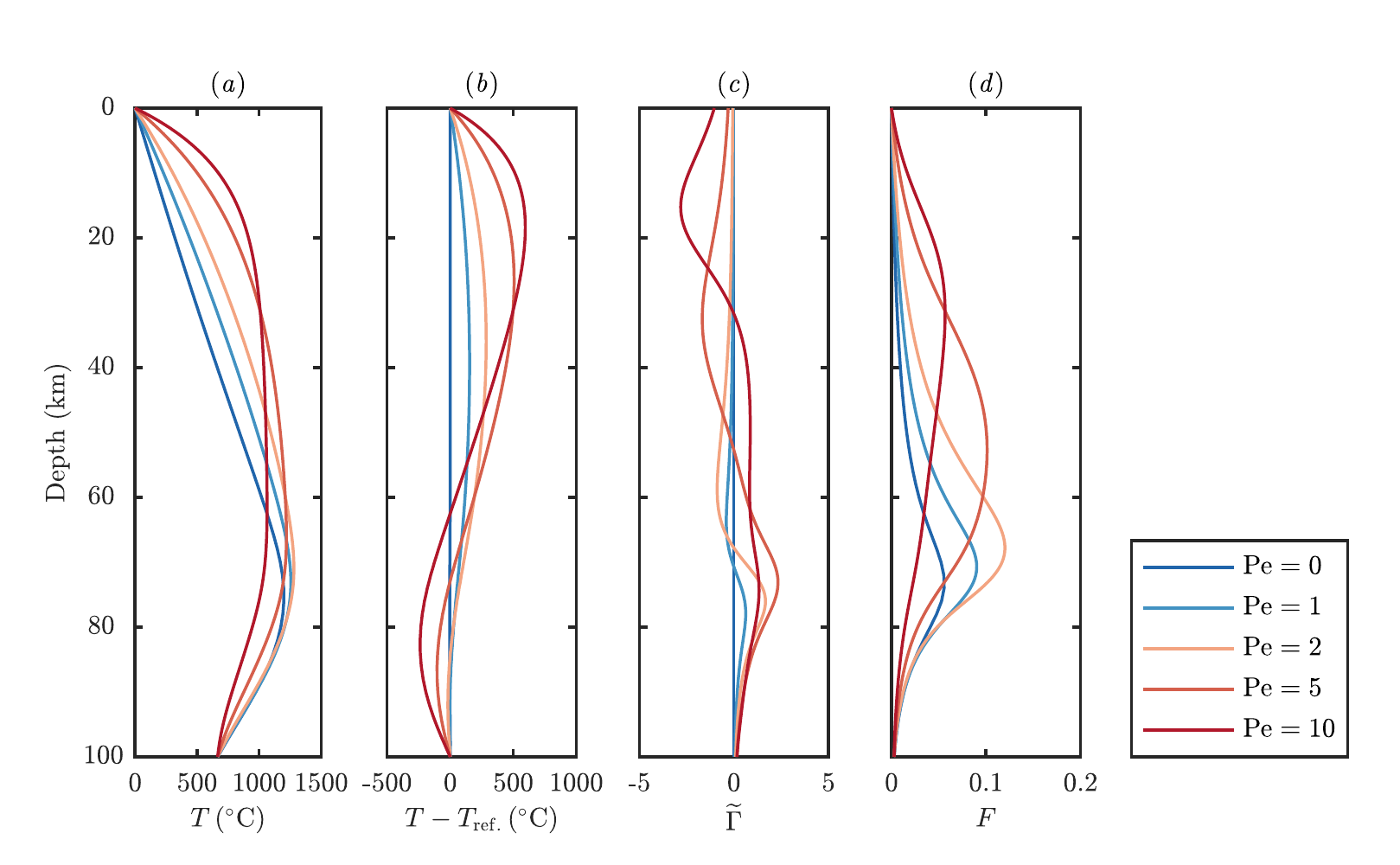}
  \caption{Melting-column model with fixed temperature at the slab at \mbox{100 km} depth and the surface. (\textit{a}) temperature profile; (\textit{b}) temperature perturbation caused by magmatism ($T-T_{\mathrm{ref.}}$); (\textit{c}) scaled melting rate $\tilde{\Gamma} = \Gamma (H^2 / \kappa \rho)$; (\textit{d}) degree of melting $F$. The range of P\'{e}clet number considered is roughly equivalent to the range of fluxes reported in Ref. \cite{Wilson2014}. Bulk water content used in the petrological model of melting is 0.5\%.} 
  \label{fig:1Dcol}
\end{figure*}

Figure~\ref{fig:1Dcol} shows results of the 1D column calculations. These are obtained for the column rising from slab where it is 100~km deep. This choice is roughly consistent with the observed mean depth beneath arc volcanoes \cite{EnglandEngdahl2004,Syracuse2006}. The flux at the base of the column is varied within the range suggested by a previous study \cite{Wilson2014}. Dimensionally, this range corresponds to fluxes between 0.2--2~m/kyr. Panel (\textit{a}) shows profiles of the absolute temperature; panel (\textit{b}) shows the temperature difference compared to the single-phase (magma-free) reference case. The change in temperature from the reference state increases with the imposed flux and is significant even at the lower end of the plausible range \cite{Wilson2014}. Immediately above the slab, upward flow reduces the mantle temperature as material is transported from the relatively cold slab. Nearer the surface, the effect is reversed as upward flow brings warm material from the mantle into the lithosphere. This effect is supplemented by latent heat associated with melting and solidification, shown in panel (\textit{c}). Above the slab, melting of the mantle wedge facilitated by the presence of water consumes latent heat. Nearer the surface, solidification of the melt deposits latent heat. The maximum degree of melting (\textit{d}) is increased because of the elevated temperatures, which will have a significant geochemical signature \cite{Turner2016}. It is interesting to note that the maximum degree of melting does not vary monotonically, but peaks at an intermediate P\'{e}clet number between 2 and 5. 

\begin{figure}[t]
\includegraphics[width=8.64cm]{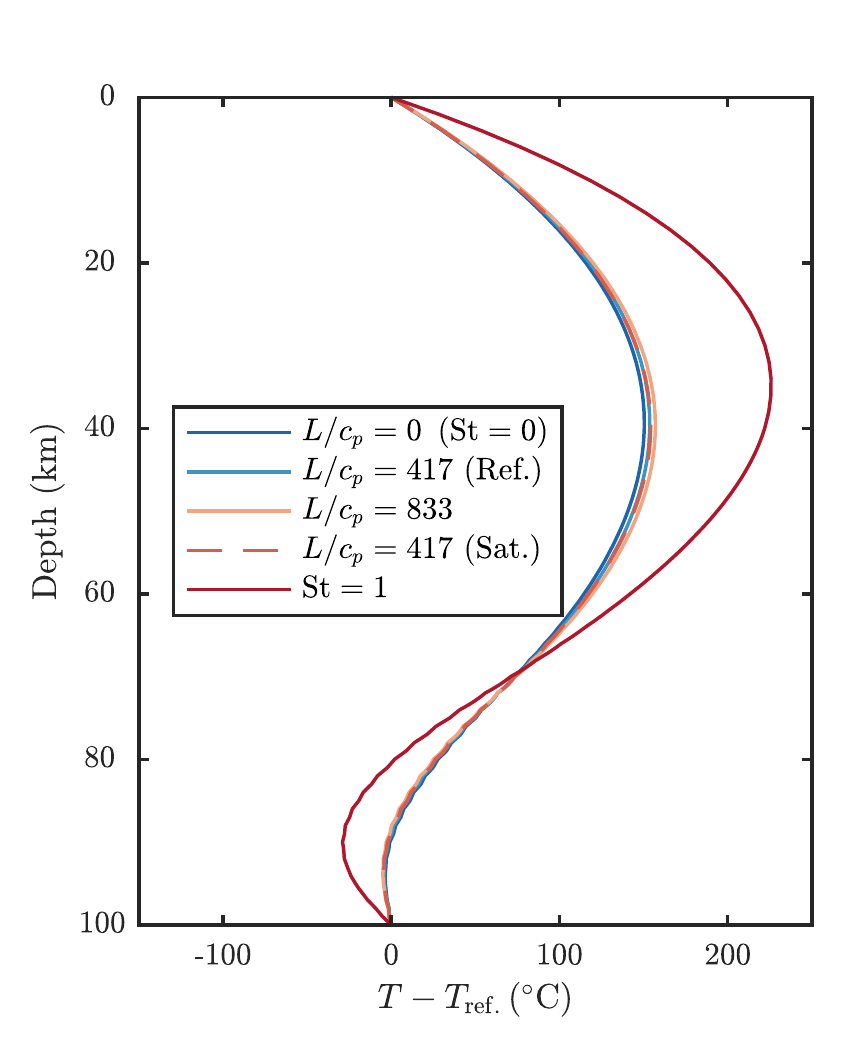}
\caption{The effect of latent heat. We show the sensitivity of the calculated thermal effect of magmatism \mbox{$T-T_\mathrm{ref.}$} to different representations of latent heat in the energy eqn.~\eqref{eq:heat1D} at fixed $\Pe=1$. We consider the case of no latent heat \mbox{($L/c_p = 0^\circ$C)} and double the reference latent heat  \mbox{($L/c_p = 833^\circ$C)}. We also consider a more detailed parameterization inspired by \citep{Katz2003} that accounts for saturation in water (\textit{cf.} Sec.~S1.3, Supplementary Material), which is labelled (Sat.). Note that the Stefan number $\St=(L/c_p)\partial F/\partial T$, and is small through the temperature range encountered, so the effect of latent heat is relatively small. We also show a calculation with a fixed Stefan number $\St=1$. In this case, the effect of latent heat is comparable to that of advection. } \label{fig:latent}
\end{figure}

The main physical mechanism giving rise to this thermal response is advection by the magma; latent heat release reinforces the advective heat flux. Additional calculations, shown in Figure~\ref{fig:latent}, demonstrate that latent heat has a subordinate effect on the temperature profiles. Other calculations shown in Fig.~\ref{fig:latent} indicate that these results are robust to changes in the parameterization of hydrous flux melting (either to mimic more closely a more detailed parameterization \cite{Katz2003}, or by arbitrarily doubling the Stefan number). The relative importance of latent to specific heat is controlled by the Stefan number $\St$. This is typically relatively small; $\St<0.1$ throughout the temperature range encountered ($T_\mathrm{ref.}\leq 1250^\circ$C, above a slab 100~km deep), as shown in Supplementary Material, Fig.~S3. If the Stefan number were much larger, latent heat release would be comparable to thermal advection by magma (Fig.~\ref{fig:latent}). A larger Stefan number may be relevant for magmatic environments dominated by melting at high isentropic productivity, above the anhydrous solidus (i.e., plumes and mid-ocean ridges). But subduction zones are characterized by low-productivity hydrous-flux melting \cite{Hirschmann1999III}, associated with a small Stefan number, and hence the role of latent heat is relatively minor. 

\subsection{Two-dimensional thermal model with magma migration}
Two-dimensional effects that are neglected in column models, such as lateral diffusion and changes to viscosity structure and mantle flow, require a more careful treatment. We next consider the thermal consequences of magmatic advection by modifying a canonical, two-dimensional reference simulation of a subduction zone \cite{vanKeken2008} to include a prescribed segregation flux $\boldsymbol{v}_D$ in the heat eqn.~\eqref{eq:heat}. We assume that magma segregates purely vertically, driven by the density difference between solid and liquid phases. We prescribe this flow in terms of Gaussian profiles centred at the typical position of the arc volcano \cite{EnglandEngdahl2004,Syracuse2006}. Our numerical scheme solves iteratively for thermal structure and solid flow, which are fully coupled through advection and the temperature dependence of mantle viscosity, until a steady state is achieved. The thermal impact of magmatism is then defined as the difference between the calculated and reference temperature fields. 

\begin{figure*}
 \includegraphics[width=16.44cm]{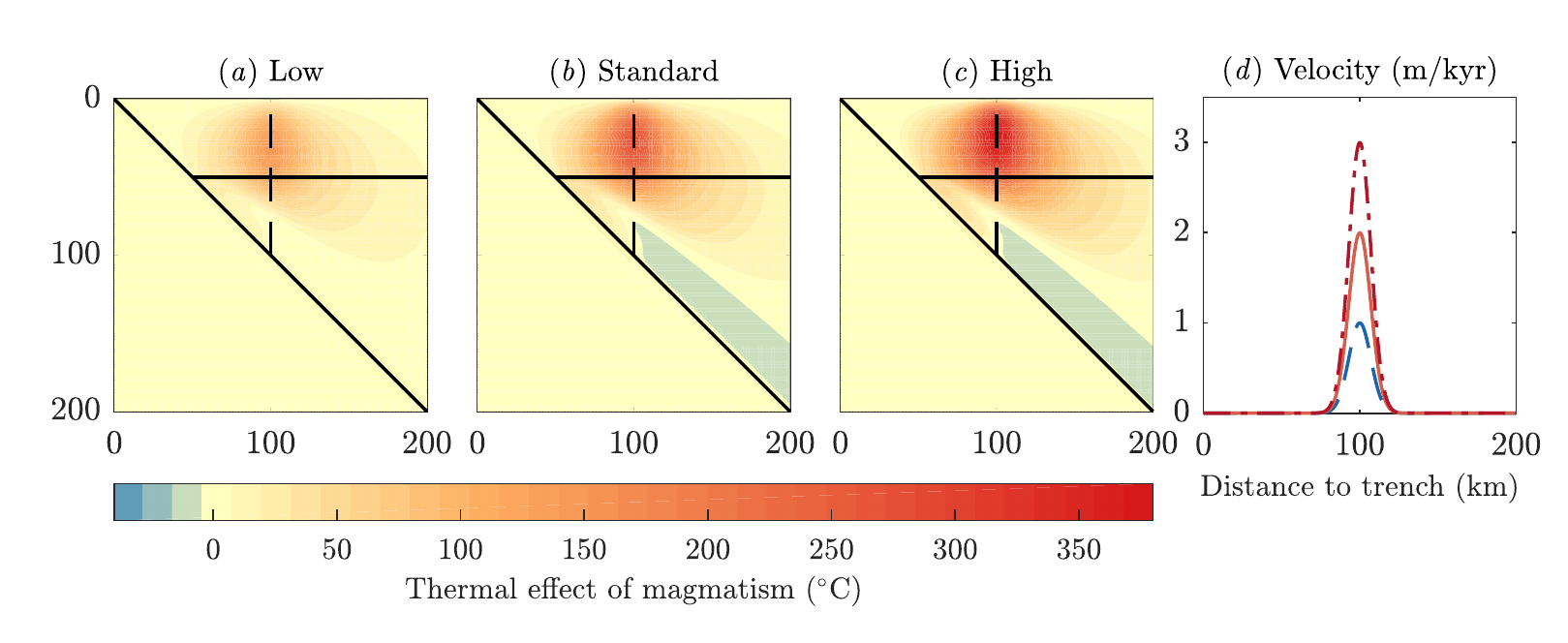}
  \caption{The thermal impact of magmatism ($T-T_{\mathrm{ref.}}$) associated with magma flow beneath the volcanic arc (dashed black line). The slab and overriding plate geometry are shown by solid black lines. We compare a low, standard, and high estimate of the magmatic flux (\textit{a}--\textit{c}). The prescribed magmatic segregation flux (vertical Darcy velocity) is shown in (\textit{d}).  Horizontal and vertical scales are distance from the trench, in kilometres.} 
  \label{fig:wedge_arc}
\end{figure*}

The two-dimensional calculations, shown in Figure~\ref{fig:wedge_arc}, predict that magmatic transport substantially alters the thermal structure in subduction zones. The main effect is to raise temperatures near the base of the lithosphere, where warm material is transported from the mantle upward. These 2D results are qualitatively consistent with the 1D column models (cooling above the slab, warming near the surface), indicating that the physical mechanisms discussed above remain pertinent. However, some features only occur in two dimensions. For example, cooling is observed immediately above the slab-top deeper than 100~km; this is caused by advection with the mantle flow. Thus the thermal impact of magmatism is distributed beyond the imposed region where the magma flows. 

Our standard estimate of the magmatic flux uses a Gaussian velocity profile (Fig.~\ref{fig:wedge_arc}\textit{d}) with a peak velocity of \mbox{2 m/kyr} and a width of \mbox{10 km}, giving a total flux comparable to global estimates \cite{Reymer1984,Crisp1984,England2010}. In this case, magmatism raises temperatures by up to 270~K (Fig.~\ref{fig:wedge_arc}\textit{b}). We also consider a magmatic flux 50\% smaller or larger than this standard case.  Temperatures are raised by $\sim$150~K (Fig.~\ref{fig:wedge_arc}\textit{a}) with the lower estimate. The higher estimate raises temperatures by up to 380~K (Fig.~\ref{fig:wedge_arc}\textit{c}). In three dimensions, the thermal effect local to arc volcanoes would likely be even greater due to along-strike flow focussing. 

\begin{figure*}
\includegraphics[width=16.44cm]{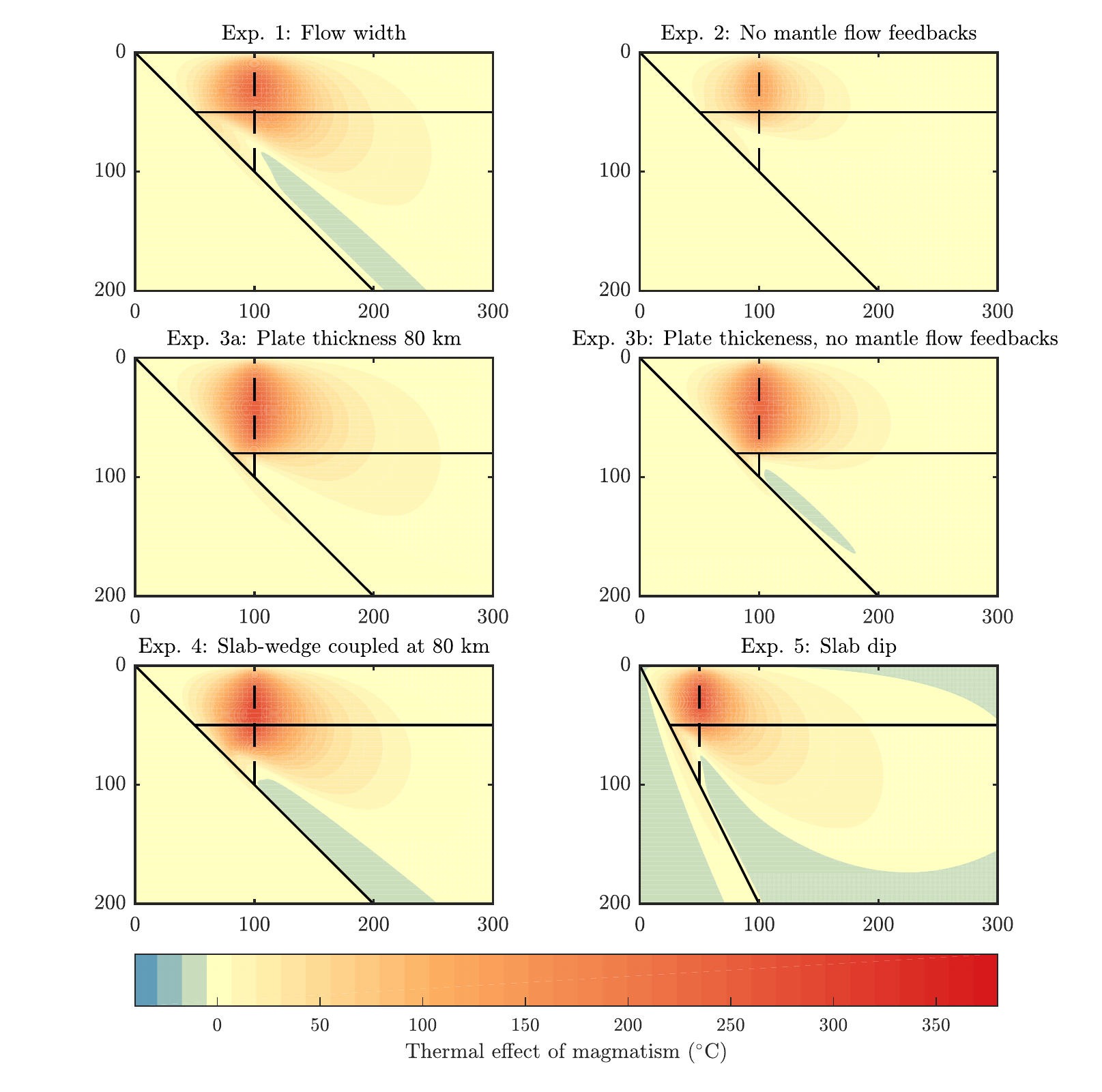}
\caption{Experiments that illustrate the sensitivity of results (Fig.~\ref{fig:wedge_arc}\textit{b}) to various modelling choices, as described in the text.} 
\label{fig:wedge_experiments}
\end{figure*}

Figure~\ref{fig:wedge_experiments} shows the results of additional calculations that explore the sensitivity to different parameter values and modelling choices that are consistent with observational constraints. For all these calculations, we compare against the standard magma flux case (Fig.~\ref{fig:wedge_arc}\textit{b}). For the impatient: these sensitivity experiments show that our key conclusion --- that magmatism has a significant thermal effect --- is robust.

First, we find that the total magma flux is more significant that the width of the flow. In Model Experiment~1, we show that similar temperatures are obtained by doubling of the width of the magma flow while halving of its magnitude to hold the total flux constant. The wider flow has a slightly lower peak (by 40~K) and is slightly more diffuse. However, these differences are minor compared to those associated with varying the total magma flux (Fig.~\ref{fig:wedge_arc}\textit{a},\textit{c}). The width of the thermal response is controlled primarily by the balance between advective heat transport by the magma and thermal diffusion.

\begin{figure*}
\includegraphics[width=12.44cm]{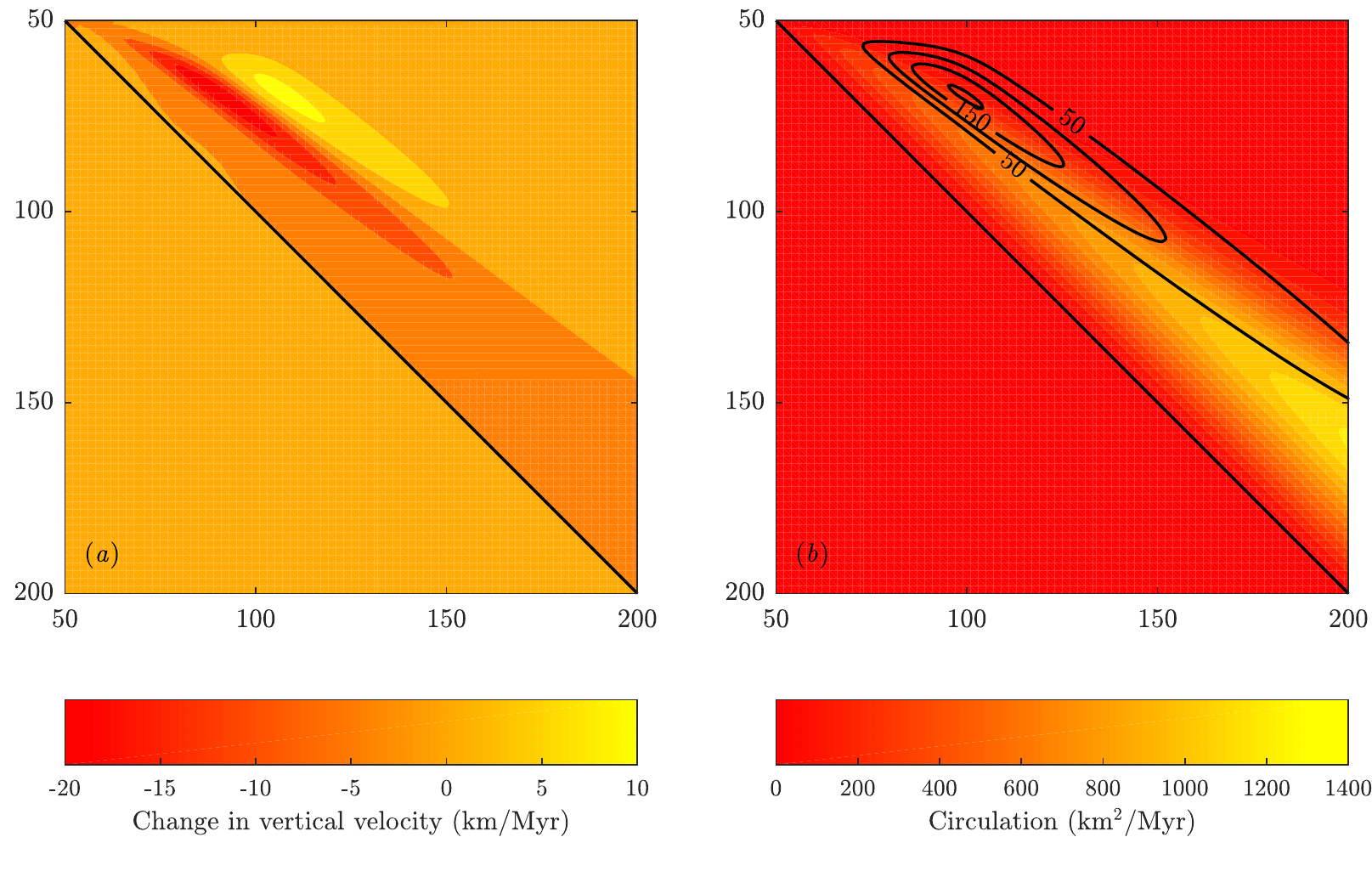}
\caption{Change in solid velocity associated with the thermal impact of magmatism. (\textit{a}) changes in vertical velocity, which are moderately significant compared to the speed of the subducting slab which is 50~km/Myr. (\textit{b}) circulation (streamfunction) is shown as the colour scale, with solid contours showing the change in the circulation due to the thermal impact of magmatism.
} \label{fig:wedge_Du}
\end{figure*}

Second, we consider the effect of the viscous coupling between the solid velocity and the temperature field. We partially decouple the model by holding the solid velocity field fixed at the reference conditions associated with the reference temperature field (i.e., that without magmatism). In Experiment~2, we show that the semi-decoupled calculations have a significantly smaller thermal response. The mechanism is as follows: in the fully coupled calculations, the elevated temperatures caused by magmatism lower the mantle viscosity, increasing the mantle wedge circulation, which is shown in Figure~\ref{fig:wedge_Du}. This leads to increased heat transport toward the arc (a positive feedback). The effect of coupling is more pronounced with smaller plate thickness because there is a larger region of mantle flow where the viscosity is reduced, leading to faster circulation (\textit{cf.}~Exps.~3a and 3b in Fig.~\ref{fig:wedge_experiments}).

Third, we consider the effect of the imposed thickness of the overiding plate (Exp.~3a of Fig.~\ref{fig:wedge_experiments}). The thermal effect of magmatism decreases slightly with increasing plate thickness. This is associated with cooler temperatures in the reference state, reducing the advection of heat by the magma. The decrease is also aided by the fact that the coupling to the solid velocity becomes a less significant positive feedback as plate thickness increases (Exp.~3b of Fig.~\ref{fig:wedge_experiments}, which is relatively similar to Exp.~3a).

Fourth, we consider the effect of slab--wedge coupling (Exp.~4 of Fig.~\ref{fig:wedge_experiments}). We increase the slab--wedge coupling depth from 50~km to 80~km, a value suggested by Ref.~\cite{Wada2009} on the basis of fore-arc heat flow measurements. This has a significant effect on the reference state without magmatism. However, it has only a small effect on the thermal effect of magmatism itself.

Fifth, we consider the effect of slab dip (Exp.~5 of Fig.~\ref{fig:wedge_experiments}). We double the slab slope from 1:1 to 2:1. Again, we find that the thermal effect of magmatism is qualitatively very similar to the standard case in Fig.~\ref{fig:wedge_arc}\textit{b}. 

\begin{figure*}
\includegraphics[width=16.44cm]{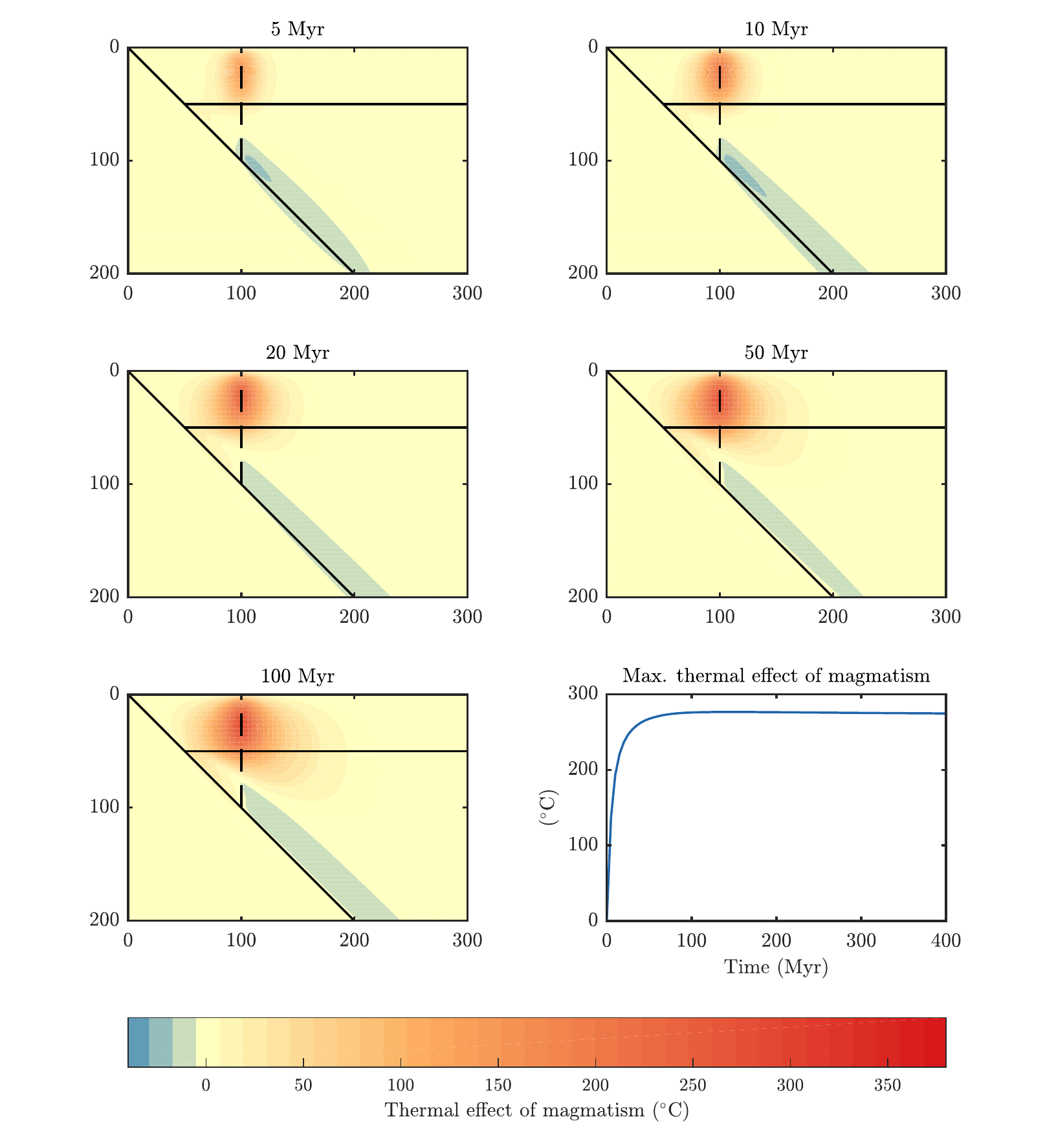}
\caption{The temporal evolution of the thermal effect of magmatism. The final panel shows the close approach to a steady state, which is achieved after around 50~Myr.} \label{fig:wedge_timeseries}
\end{figure*}

Finally, in Figure~\ref{fig:wedge_timeseries}, we consider the transient evolution towards steady state. We use an initial condition corresponding to old oceanic lithosphere and impose the same fixed magma flux. The thermal effect of magmatism evolves to a steady state over a period of about 50~Myr, controlled by thermal diffusion, although the thermal structure much further away from the arc evolves on a longer timescale \cite{Hall2012}. The transient spatial pattern of elevated sub-arc temperatures is consistent with the steady-state pattern. However, the magnitude of the thermal effect depends on the age of the subduction zone. 

In the Supplementary Material, Sec.~S3, we consider separately the magmatism associated with each of the major slab dehydration reactions that occur at various depths. 

In summary, in each sensitivity test, we find that although small quantitative differences in the results are produced, the overall behaviour and the basic conclusion is similar. Thus the thermal effect of magmatism we show in Figure~\ref{fig:wedge_arc} is robust; the details will vary between subduction zones, but the physical effect is to significantly modify the thermal structure from that predicted by canonical models.

\section{Discussion and Conclusions}
\begin{figure} 
\includegraphics[width=8.64cm]{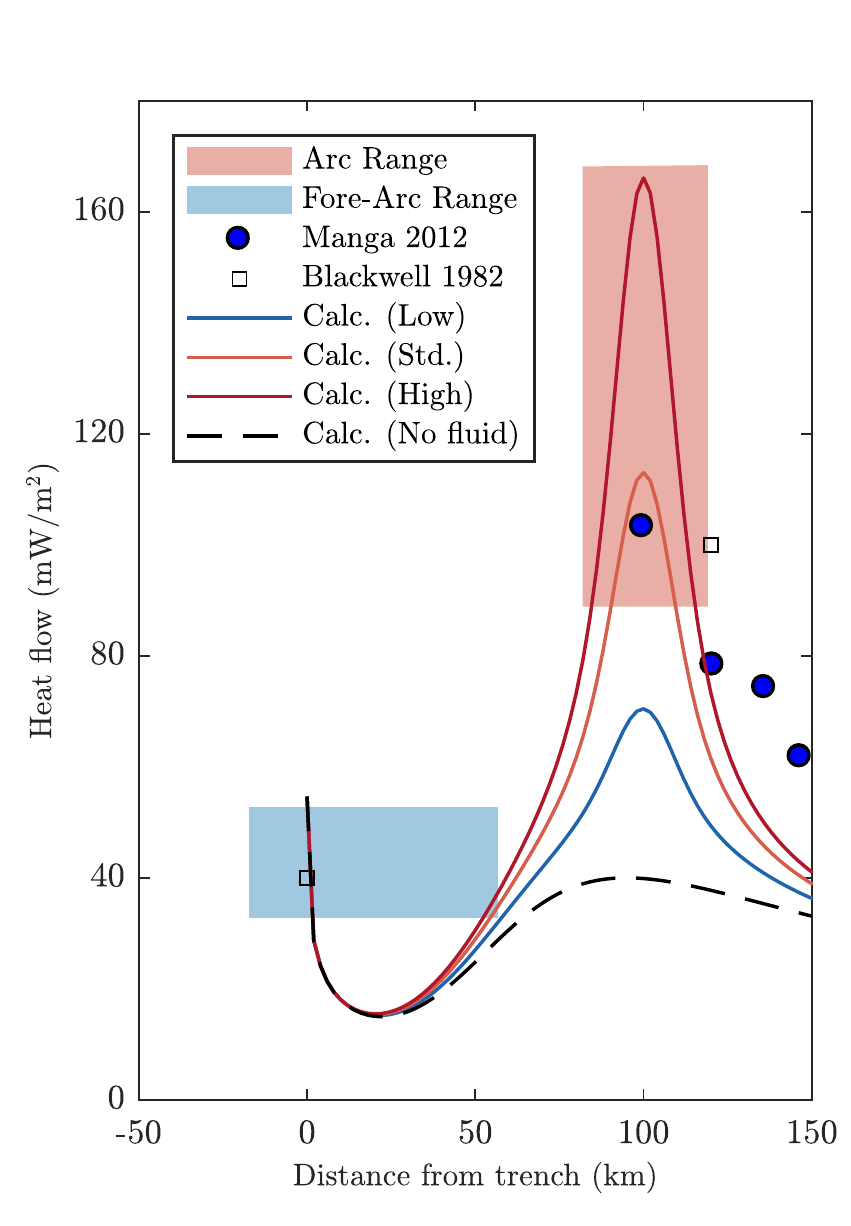}
  \caption{Predicted arc heat flow in subduction zones associated with melt migration compared with observed, global ranges. The ranges shown are based on the global compilation of \cite{Stein2003}, as presented by \cite{Manga2012}. Also plotted are local measurements from  oceanic \cite{Manga2012} and continental \cite{Blackwell1982} subduction zones. The heat flow is raised by around 40--120~\mbox{mW/m$^2$}, concentrated near the region of peak magma flow, 100~km from the trench. Model results were obtained by evaluating surface temperature gradients in calculations shown in Fig.~\ref{fig:wedge_arc} and converting to heat flow using a constant thermal conductivity of 2.52~\mbox{W/m/K}.} 
  \label{fig:wedge_heatflow_comb}
\end{figure}

Our results are consistent with heat flow and petrological observations. The elevated heat flow measured in subduction zones, shown in Figure~\ref{fig:wedge_heatflow_comb}, can be associated with elevated near-surface temperatures. This elevated heat flow is strongest at the position of the arc, over a width of around 50~km. The width is determined by thermal diffusion rather than the imposed width of magma flow. Our models that use a magma flux between the standard and high values are consistent with heat-flow observations near the volcanic arc. Note that the low fore-arc heat flow in our models is an artefact of the simplified geometry, particularly the constant slab dip. Furthermore, hydrothermal circulation in the subducting crust has a significant thermal effect in the fore-arc region, consistent with heat flow observations along the Chilean subduction zone \cite{Spinelli2016}.  Similarly, we find that magmatic flow has a significant thermal effect in the sub-arc region, consistent with heat flow observations there.

\begin{figure}
\includegraphics[width=8.64cm]{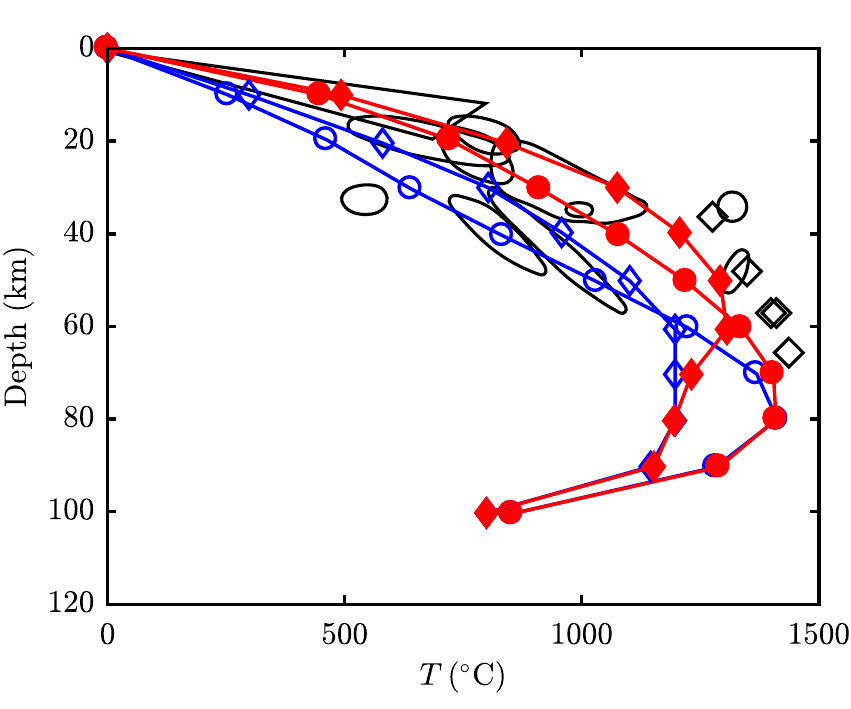}
  \caption{Temperature structure compared to a compilation of petrological and heat flow data (black open shapes are taken from Plate~1 in Reference~\cite{Kelemen2003}). The output of two thermal models \cite{Furukawa1993,vanKeken2002} are temperature-shifted by the thermal impact of melt migration, calculated as the standard case in Fig.~\ref{fig:wedge_arc}(\textit{b}). This shift is most sensitive to the total magmatic flux. The original model temperatures are open blue circles and diamonds; the shifted temperatures are shown in solid red markers of the corresponding shape.} 
  \label{fig:Kelemen_compilation}
\end{figure}

Evidence from petrological observations in Figure~\ref{fig:Kelemen_compilation} suggests that temperatures in subduction zones are some 200--300~K hotter than would be expected on the basis of canonical models of mantle flow alone \cite{Kelemen2003,Perrin2016}. This discrepancy peaks at around 60~km depth, comparable to the depth where we find magmatism has the greatest thermal impact. Inclusion of melt migration in thermal models can reconcile much of this discrepancy. This consistency between observation and thermal modelling supports the hypothesis that magmatism significantly alters the thermal structure of subduction zones. 

Scaling arguments also support our hypothesis. Indeed, it is possible to approximate the effect on heat flow due to magmatic advection as follows. The elevated heat flow is 
\begin{equation}
  Q \approx \frac{F_V \rho c_p \Delta T  }{A} \approx 80 \, \mathrm{mW /m}^{2}, 
\end{equation} 
based on a global magma flux \mbox{$F_V=1$}~\mbox{km$^3$ yr$^{-1}$} \cite{Reymer1984}, density \mbox{$\rho=3\times 10^{3}$}~\mbox{kg m$^{-3}$}, heat capacity \mbox{$c_p=1.2\times 10^{3}$} \mbox{J kg$^{-1}$ K$^{-1}$}, $\Delta T \approx 1350$~K, and an area of elevated heat flow \mbox{$A\sim2\times 10^{12}$}~\mbox{m$^2$} (the total length of $50 \times 10^3$~km and an assumed width of 40~km). This is consistent with Fig.~\ref{fig:wedge_heatflow_comb}. 

We can also estimate the ratio $R$ of advective heat transport by magma to the latent heat release (the two mechanisms by which magmatism changes the thermal structure):
\begin{equation}
  R \approx  \frac{\rho c_p \left| \boldsymbol{v}_D \right|  (\Delta T/H)}{L\Gamma}  \approx \frac{c_p \Delta T}{L} \frac{\rho \left| \boldsymbol{v}_D \right| }{\Gamma H} \approx \frac{c_p \Delta T}{L} \approx 3.2,
\end{equation} 
where \mbox{$L=5\times 10^{5}$ J kg$^{-1}$}. We used the fact that ${\rho \left| \boldsymbol{v}_D \right|} /{\Gamma H} \approx 1$ on average at steady state, since there is a balance between melt production, melt extraction, and melt solidification. Therefore, magmatism has a significant thermal effect and this effect is mainly due to advection by the magma. This latter finding is in contrast to a previous, simpler, one-dimensional model \cite{England2010,Perrin2016}.

The thermal signature of melt migration should be considered when interpreting heat flow, petrologic, gravity, and seismic data. Seismic velocities and attenuation depend strongly on temperature \cite{Takei2017}. Thus our results suggest that a part of the measured low seismic velocities and high attenuation beneath the arc is likely associated with high temperatures. However, the relatively small spatial extent of the thermal anomalies we predict ($\sim$50--100~km) will make them difficult to observe seismically. A perturbation as large as 300~K also increases the maximum degree of melting, which in turn affects the chemistry of arc volcanoes (or our inferences about the mantle made on the basis of geochemical measurements) \cite{Turner2016}. It also significantly affects the solid mantle flow through reduction of mantle viscosity, leading to increased circulation in the mantle wedge \cite{vanKeken2002}. Furthermore, thermal structure affects magma pathways in subduction zones, focussing magmas along the thermal lithosphere from a broader area to beneath the arc volcanoes \cite{Sparks1991, Wilson2014}. Thus, coupled mantle--magma flow may well affect the location of arc volcanoes themselves, consistent with evidence from global systematics \cite{England2010}.

\section*{Author contributions} R.F.K.~conceived the study. D.R.J.~and M.T. developed the one-dimensional melting column model and petrological model of melting. D.R.J., R.F.K.~and J.F.R. developed the two-dimensional thermal model with magmatism. J.F.R. contributed a single-phase numerical code to compute the thermal structure of a subduction zone, to which D.R.J. added two-phase flow. R.F.K.~and D.R.J. compared the model with petrological and heat flow observations. D.R.J. wrote the manuscript with R.F.K., and discussed the manuscript with M.T.~and J.F.R. All authors jointly discussed and analyzed the data, results, conclusions, and implications.

\section*{Acknowledgements} The authors thank  D.~McKenzie, P.~England, B.~Hacker and J.~Ague for comments on an earlier version of this manuscript. We would like to thank the Isaac Newton Institute for Mathematical Sciences for its hospitality during the programme Melt in the Mantle which was supported by EPSRC Grant Number EP/K032208/1. D.R.J.~acknowledges research funding through the NERC Consortium grant NE/M000427/1 and NERC Standard grant NE/I026995/1. The research of R.F.K.~leading to these results has received funding from the European Research Council under the European Union's Seventh Framework Programme (FP7/2007--2013)/ERC grant agreement number 279925. M.T.~received research funding from the Royal Society Newton International Fellowship. J.F.R.~thanks the Leverhulme Trust for support. The authors would also like to thank the Deep Carbon Observatory of the Sloan Foundation.

\noindent {\centering \textbf{{\large{SUPPLEMENTARY MATERIAL}} } }
\setcounter{section}{0}
\setcounter{figure}{0}
\setcounter{equation}{0}

\vspace{0.5cm}
\textbf{
\noindent The Supplementary Material contains further details of the petrological model of melting (Sec.~\ref{sec:petrological_model}), the 1D column model (Sec.~\ref{sec:1Dcol}), and the 2D thermal model (Sec.~\ref{sec:2D_details}).
}

\renewcommand\thefigure{S\arabic{figure}}  
\renewcommand\thesection{S\arabic{section}}  
\renewcommand\theequation{S\arabic{equation}}

\section{Petrological model of hydrous flux melting} \label{sec:petrological_model}
In the 1D column model, we use a simple petrological model of hydrous flux melting, which is the dominant form of melting in subduction zones. The model was inspired by previous studies \cite{Hirschmann1999III,Katz2003,Keller2016}, and is developed as follows. First, we restrict attention to a ternary system. The three components should not be thought of as identifiable minerals or oxides but rather as idealized components chosen to capture the physics in which we are interested. We start with two components that can be considered `refractory' and `fertile' \cite{Ribe1985,Hewitt2010}. To this system, we add a third component to represent volatiles. We initially take this component to be `water' and we consider that the concentration of `water' is relatively small. One role of this third hydrous component is to depress the solidus temperature.

Our second simplification is to use a quasi-linear phase diagram. This can be thought of as a linearization of the ternary phase loops used by Ref.~\cite{Keller2016} about some initial composition. 

Our third simplification is that the melting/solidification reactions happen sufficiently rapidly that a partially molten region is at local thermodynamic equilibrium. This implies that compositions of the coexisting solid and liquid phases are given exactly by the phase diagram. 

\subsection{Mathematical description of phase diagram}
The solidus temperature increases with increasing pressure at a rate $\gamma$.  We linearize the dependence of the solidus on chemical composition. Since the sum of the concentrations of the components is unity, we need only specify two linear coefficients $M_2, M_3$ for the fertile and water components respectively, both of which lower the solidus temperature. Thus the solidus temperature
\setcounter{equation}{4}
\begin{equation} \label{eq:solidus}
T_s=T_{s0}-\rho g z /\gamma - M_2 c^s_2 - M_3 c^s_3.
\end{equation}
This expression can be rearranged to give, for example, the solidus concentration $c^s_2$ as a function of temperature, depth, and concentration of the third component. An interpretation of equation~\eqref{eq:solidus} can be made by identifying $T_{s0}-\rho g z /\gamma$ with the solidus temperature of the refractory component at given depth $z$, which in this section is negative. 

We assume that the liquidus concentration is related to the solidus concentration as follows:
\begin{align} 
c^l_2 &= c^s_2+\Delta c_2, \label{eq:liquidus2} \\
c^l_3 &= c^s_3+\Delta c_3. \label{eq:liquidus3}
\end{align}
For the simplest case we take $\Delta c_{2,3}$ to be constants, but we will also consider generalizations.

\subsection{Choice of parameter values and implications for melting}
We choose parameters in our model to constrain the degree of so-called `batch melting'  as a function of temperature and pressure:
\begin{equation} \label{eq:F_comp}
F= \frac{ \widetilde {c_{0j}}-c_j^s}{c^l_j-c^s_j}.
\end{equation}
Batch melting refers to the degree of melting experienced by a sample raised to given temperature and pressure conditions assuming no extraction of melt. The composition $\widetilde {c_{0j}}$ is the composition of the solid mantle before the onset of melting. We then combine equations \eqref{eq:solidus}--\eqref{eq:liquidus3}, which apply for each $j$, with equation \eqref{eq:F_comp} to obtain 
\begin{equation}
F= \frac{\widetilde{c_{03}}-c_3^s}{\Delta c_3}=\frac{T-T_{s0}+\rho g z/\gamma+M_2 \widetilde {c_{02}} +M_3 \widetilde {c_{03}}}{M_2 \Delta c_2 + M_3 \Delta c_3}.
\end{equation}
A key quantity is the isobaric productivity $\partial F/\partial T$. If $\Delta c_2$ and $\Delta c_3$ are constants, then the isobaric productivity is a constant
\begin{equation}
\frac{\partial F}{\partial T} = \frac{1}{M_2 \Delta c_2 + M_3 \Delta c_3}.
\end{equation}
Thus melt is produced at a constant rate with increasing temperature. Linear models of two component melting already include this effect \citep[e.g.][]{Hewitt2010}. It is worth noting that the productivity is reduced by the third, hydrous component.

In this formulation, volatiles do indeed depress the solidus temperature. However, in addition to depressing the solidus, volatiles are also associated with a `low-productivity tail' \cite{Hirschmann1999III}. The initial melting above the solidus temperature is less productive than later melting: 
\begin{equation}
\left. \frac{\partial F}{\partial T} \right|_{F=0} <  \left. \frac{\partial F}{\partial T} \right|_{F=1}.
\end{equation}
The purely linear model does not satisfy this constraint, because the productivity is constant. Therefore, we generalize our model to allow for a low-productivity tail. Volatiles are incompatible and partition into the melt with a partitioning coefficient $D$ defined by
\begin{equation} \label{eq:partition}
c_3^s=Dc_3^l \Rightarrow \Delta c_3= c_3^s(1/D-1),
\end{equation}
where $D\ll1$ for incompatible, volatile elements. We assume that $D$ is constant. However, $\Delta c_3$ is no longer constant, instead depending on composition, and hence pressure and temperature. Upon a little rearrangement, we find
\begin{align}
T-&(T_{s0}-\rho g z /\gamma - M_2 \widetilde{c_{02}} - M_3 \widetilde{c_{03}}) \nonumber \\ &=(M_2 \Delta c_2 / \Delta c_3+M_3) ( \widetilde{c_{03}}-c_3^s),
\end{align}
which can be rearranged to give a quadratic equation for $c_3^s$, recalling that $\Delta c_3$ is proportional to $c_3^s$. The degree of melting $F$ is no longer a linear function (however it can be computed explicitly using the quadratic formula so there is no computational difficulty, unlike more complex nonlinearities where iterative methods are required to solve for $F$). We can calculate the isobaric productivity at $F=0$ and $F=1$ and find
\begin{align}
\left. \frac{\partial F}{\partial T} \right|_{F=0} &= \left[ M_3 \widetilde{c_{03}}(1/D-1) +M_2 \Delta c_2\right]^{-1}, \label{eq:prodF0} \\
\left. \frac{\partial F}{\partial T} \right|_{F=1} &= \left[ M_3 \widetilde{c_{03}}D(1-D) +M_2 \Delta c_2\right]^{-1}. \label{eq:prodF1} 
\end{align}
We can interpret the effective isobaric productivity of the mixture as the harmonic mean of productivities associated with the fertile and volatile component. Typically, the contribution of the volatile component dominates at small $F$ and the fertile component dominates at large $F$ because 
\begin{equation}
M_3 \widetilde{c_{03}}D(1-D)  \ll M_2 \Delta c_2 \ll M_3 \widetilde{c_{03}}(1/D-1).
\end{equation}
Our formulation thus achieves the low-productivity tail expected physically; and it recovers the fertile--refractory system in the absence of volatiles.

To summarize, the degree of melting increases over the temperature range 
\begin{align}
T(F=0) &= T_{s0}-\rho g z /\gamma - M_2  \widetilde{c_{02}} - M_3  \widetilde{c_{03}}, \label{eq:TF0} \\
T(F=1) &= T_{s0}-\rho g z /\gamma - M_2  \widetilde{c_{02}} + M_2 \Delta c_2 - M_3  D \widetilde{c_{03}} \label{eq:TF1}.
\end{align}
The gradient of the function $F(T)$ at these endpoints is given by equations~\eqref{eq:prodF0} and \eqref{eq:prodF1}. 

\begin{figure}[t]
\includegraphics[width=8.64cm]{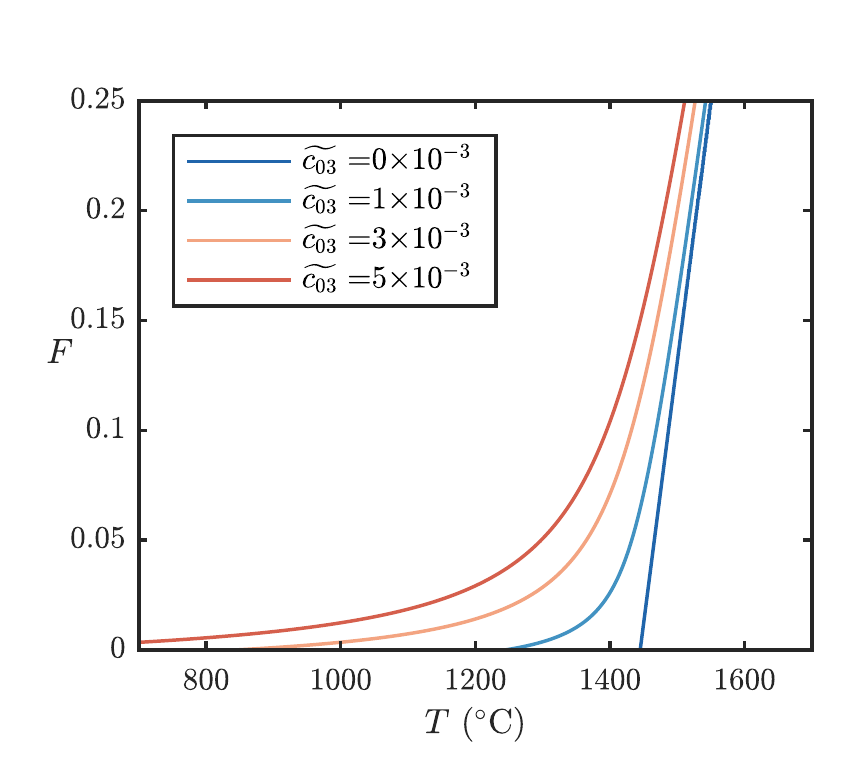}
\caption{The degree of melting $F$ as a function of temperature $T$ at increasing water concentration $\widetilde{c_{03}}$. Other parameters were fixed, namely \mbox{$T_{s0}-\rho g z/\gamma = 1550$ $^\circ$C}  at \mbox{$z=100$ km}, \mbox{$M_2=700^\circ$C}, \mbox{$M_3=2\times10^5$ $^\circ$C}, $D=0.01$, $\Delta c_2=0.6$, and $\widetilde{c_{02}}=0.15$. These parameter values are motivated by Ref.~\cite{Katz2003}.} \label{fig:degreeF(T)}
\end{figure}

Thus a limited number of parameters can describe a significant range of realistic melting behaviours, as shown in Fig.~\ref{fig:degreeF(T)}. For the anhydrous, fertile--refractory part of the system, we use \mbox{$T_{s0} = 1100$ $^\circ$C},  $\rho g/\gamma = 4.5\times10^{-3}$ \mbox{$^\circ$C/m}, \mbox{$M_2=700^\circ$C}, $\Delta c_2=0.6$, and $\widetilde{c_{02}}=0.15$. This ensures that we match the anhydrous melting curve of Ref.~\cite{Katz2003}, particularly around 3~GPa. For the hydrous part of the system, we use \mbox{$M_3=2\times10^5$ $^\circ$C}, $D=0.01$. These  parameter values were chosen to roughly match the hydrous melting curve of Ref.~\cite{Katz2003}, particularly around 3~GPa with 0.5~wt\% water. Although the precise parameter values are in the right region for consistency with previous studies and their associated experimental libraries, as well as inferences from field observations, this parameterization is too simple to reproduce all the features observed experimentally. However, it can reproduce the two main features: solidus depression and a low-productivity tail.

\subsection{Generalized model: accounting for saturation in water} \label{sec:saturation}
\begin{figure}[t]
\includegraphics[width=8.64cm]{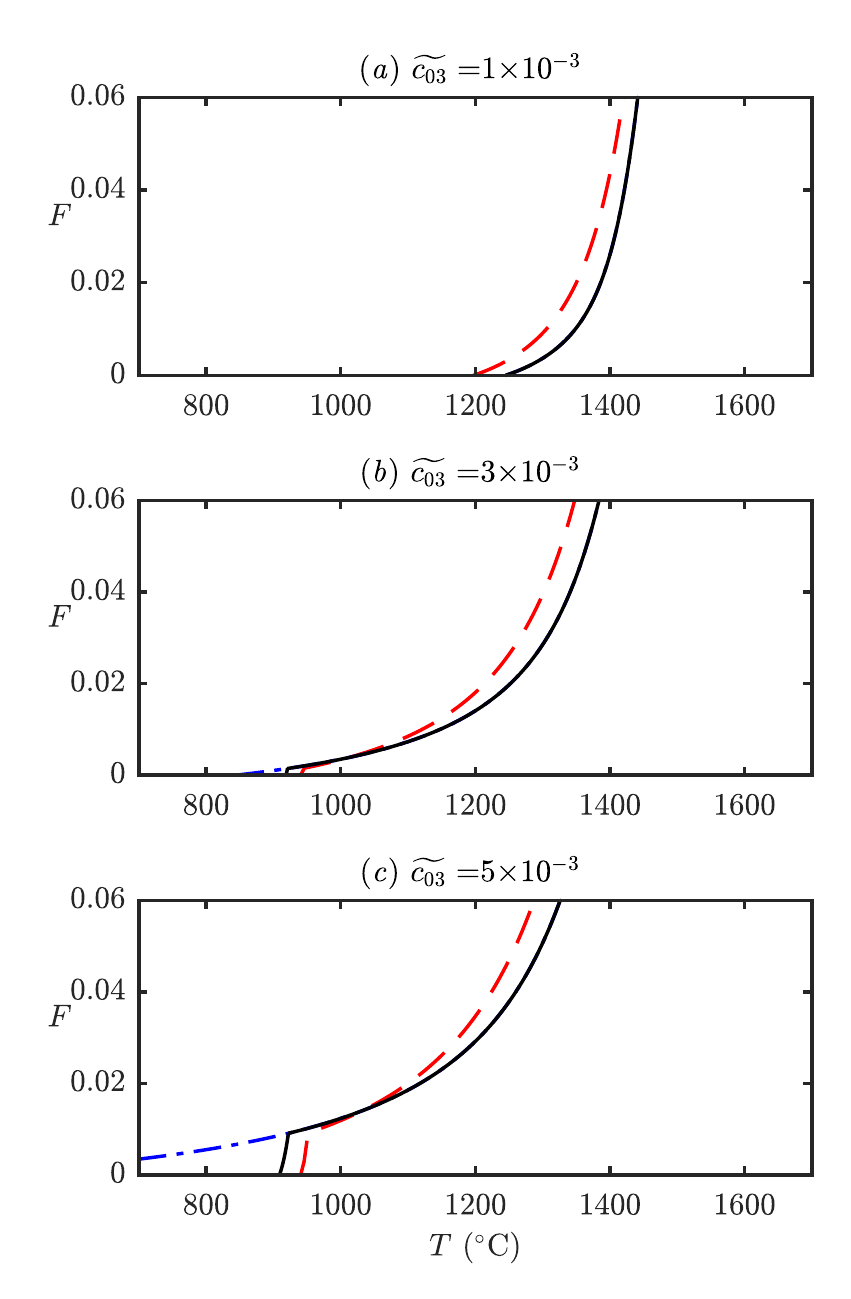}
\caption{The degree of melting $F(T)$ produced by our revised model (solid black) and the parameterization of Ref.~\cite{Katz2003} (dashed red). Results computed at fixed pressure (3~GPa, corresponding to \mbox{$z=100$ km}) at increasing water content: (\textit{a}) $\widetilde{c_{03}}=1\times10^{-3}$, (\textit{b})  $\widetilde{c_{03}}=3 \times 10^{-3}$, and (\textit{c}) $\widetilde{c_{03}}=5 \times 10^{-3}$. Note the kink in the curves around \mbox{$950^\circ$C} in (\textit{b}, \textit{c}), which is associated with water saturation. Without this saturation behaviour, our standard model predicts melting at several hundred degrees cooler temperatures (dashed blue curves).} \label{fig:Katz-degreeF-comp}
\end{figure}

The addition of more water does not indefinitely lower the solidus, because eventually water becomes saturated in the liquid phase. The amount of water that dissolves increases with pressure; Ref.~\cite{Katz2003} uses the formula
\begin{equation} \label{eq:Xsat}
X^\mathrm{sat}_\mathrm{H_20}=  12.00 P^{0.6}+1.00P,
\end{equation}
where the pressure $P$ is measured in GPa. This is well constrained by experiment below 2~GPa, and constrained indirectly at higher pressures. This corresponds to a critical degree of melting and critical temperature below which the degree of melting drops rapidly to zero, as shown in  Fig.~\ref{fig:Katz-degreeF-comp}\textit{c}, for example.

Our modelling approach is to mimic this behaviour by modifying the phase diagram. We first compute the corresponding critical solid saturation point $ c^s_\mathrm{sat}$, using equation~\eqref{eq:Xsat} for the liquid saturation and the partition coefficient of equation~\eqref{eq:partition}. For temperatures below this point, we change the freezing point depression coefficient:
\begin{equation} \label{eq:solidus_mod}
T_s=T_{s0}-\rho g z /\gamma - M_2 c^s_2 -M_4(c^s_3 - c^s_\mathrm{sat}) - M_3 c^s_\mathrm{sat},
\end{equation}
where $M_4 \leq M_3$. Note that the previous model is a special case $M_4=M_3$, and a eutectic-like phase diagram can be obtained by the special case $M_4=0$. In practice, we find $M_4=M_3/50$ makes a decent approximation to Ref.~\cite{Katz2003}, as shown in Fig.~\ref{fig:Katz-degreeF-comp}. This means that the initial productivity near $F=0$ is a factor $M_3/M_4 = 50$ times greater. We use this generalized model to assess the significance of the increased productivity near water saturation in Fig.~3 of the main text. 

%\clearpage

\section{Further details of one-dimensional column model} \label{sec:1Dcol}
In the context of a one-dimensional melting model, mass conservation imposes a strong constraint on the model behaviour in steady state. We adopt an extended Boussinesq approximation in which density differences between the phases are neglected except for their role in driving buoyant liquid segregation.  There are several equivalent ways to present the following equations; we approach the problem by considering conservation in the liquid phase and in the two-phase composite. 

Mass conservation gives
\begin{align}
\frac{d}{dz} ({\phi w^l}) & =\frac{ \Gamma}{\rho}, \label{eq:mass_cons} \\
\frac{d}{dz} \overline{w} &=0, \label{eq:bulkmass}
\end{align}
where $\overline{ x}=x^s (1-\phi) +  x^l \phi$ denotes an average over the solid and liquid phases, with volume fractions $(1-\phi)$ and $\phi$ respectively. The vertical velocity is $w$, volumetric melting rate is $\Gamma$ and density is $\rho$. We first integrate equation \eqref{eq:bulkmass} to obtain 
\begin{equation} \label{eq:F_def}
\frac{\phi w^l}{\overline{W_0}} + \frac{(1-\phi) w^s}{\overline{W_0}}=1,
\end{equation}
where ${\overline{W_0}}$ is the total volume flux at the bottom of the melting column (which is not the motion of the solid phase alone, unlike in upwelling mantle columns used in the context of mid-ocean ridge magmatism). We follow the approach of Ref.~\cite{Ribe1985} and define the quantity $F={\phi w^l}/{\overline{W_0}}$. Thus the scaled, liquid-phase volume flux is $F$ and the scaled solid phase volume flux is $(1-F)$. 

We can recover our previous definition of $F$ in equation~\eqref{eq:F_comp} by considering conservation of species mass. For each component $j=1,2,3$,
\begin{align}
\frac{d}{dz} ({\phi w^l c^l_j}) & =\frac{ \Gamma_j}{\rho}, \label{eq:species_cons}\\
\frac{d}{dz} \overline{w c_j} &=0. \label{eq:bulkspecies}
\end{align} 
Note that, by summing equation \eqref{eq:species_cons} over $j$ and comparing with equation \eqref{eq:mass_cons}, $\sum_j \Gamma_j=\Gamma$. We integrate equation~\eqref{eq:bulkspecies} and use equation~\eqref{eq:F_def} to obtain
\begin{equation} \label{eq:Fcomp_Ribe}
Fc^l_j +(1-F)c^s_j= \frac{\overline{W_0 c_{0j}}}{\overline{W_0}} \equiv \widetilde {c_{0j}}.
\end{equation}
 
We then determine the degree of melting $F$, which is controlled by an energy equation and our phase diagram. One unusual feature of subduction zones is the non-monotonic temperature profile, which is largely controlled by the flow of the solid mantle. As described in the main text, we use a steady energy balance for a one-dimensional column
\begin{equation}  \label{eq:1Dheat}
\rho c_p \overline{W_0} \frac{d T}{dz}  = -L \Gamma + \frac{d }{dz}\left(\rho c_p \kappa \frac{d T}{dz}\right) + \rho c_p \Psi, 
\end{equation}
with a volumetric source term $\rho c_p\Psi$ that represents the heat supplied by large-scale mantle corner flow. In the absence of melting, the final pair of terms on the right-hand side  establishes a non-monotonic temperature profile. Note that we do not solve momentum equations because only the two-phase average velocity $\overline{W_0}$, which is constant as a result of mass conservation, appears in the heat equation \eqref{eq:1Dheat}.

\begin{figure}[t]
\includegraphics[width=8.64cm]{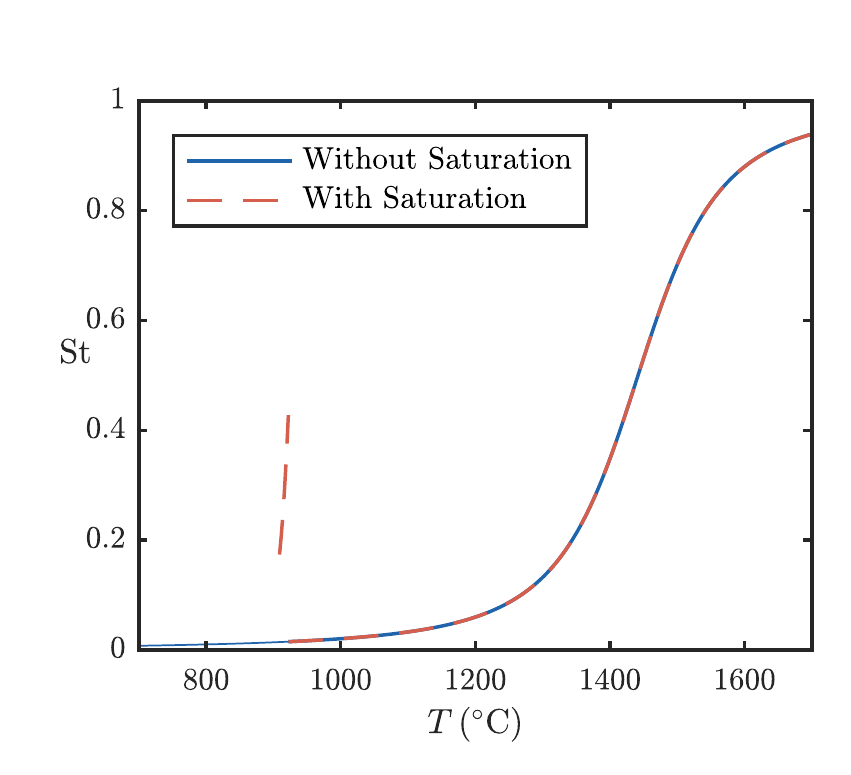}
\caption{Stefan number $\St=(L/c_p)\partial F/\partial T$ at $P=3$~GPa with and without modified phase diagram to account for water saturation, as discussed in section \ref{sec:saturation}. Note that there is now an interval of higher isobaric productivity between the solidus temperature and the temperature at which the melt ceases to be saturated.} \label{fig:stefan}
\end{figure}

Next we observe that $\overline{W_0}F'=\Gamma/\rho$ and $F'$ is proportional to the isobaric productivity discussed previously, namely $F'=(T'+\rho g /\gamma) \partial F/\partial T$. We can better understand the system by rescaling the energy equation. We scale lengths by $H$ (the depth of the melting column), and the source term by $\kappa/H^2$. The dimensionless parameters involved are a P\'{e}clet number $\Pe=H\overline{W_0}/ \kappa$, a Stefan number $\St=(L/c_p)\partial F/\partial T$, a temperature change $\Delta T_H= \rho g H/\gamma$. Then the energy equation is 
\begin{equation} \label{eq:energy}
T'' =-\Psi + \Pe\left[T'(1+\St)+\St \Delta T_H \right], \qquad (0 \leq z \leq 1).
\end{equation}
A scaled version of the melting rate is
\begin{equation}
\widetilde{\Gamma} \equiv \frac{H^2}{\kappa} \frac{\Gamma }{\rho } = \Pe \frac{\partial F}{\partial T} (T' + \Delta T_H),
\end{equation}  
which has units of degrees Kelvin. Equation~\eqref{eq:energy} is subject to boundary conditions on $T$ at $z=0$ and $z=1$. In general, the P\'{e}clet number is fixed but the Stefan number depends on temperature and pressure (hence depth), as well as the compositional parameters of our melting model. We plot the Stefan number in Fig.~\ref{fig:stefan}.

Our column-model approach is as follows. Extract a vertical temperature profile $T_{\mathrm{ref.}}$ from a single-phase mantle flow and thermal model of a subduction zone, as shown in the main article. We then calculate $\Psi=-T''_{\mathrm{ref.}}(z)$. To investigate the effect of melting, we solve the rescaled energy equation, focussing on the effect of varying the P\'{e}clet number and Stefan number (since the decompression term $\Delta T_H$ is well known). 
We present results in the main article.

%\clearpage 
\section{Further details of two-dimensional thermal model} \label{sec:2D_details}
Sources of fluids in subduction zones that trigger silicic magmatism are believed to be localized to particular depth ranges, associated with particular dehydration reactions in the subducting slab. Thus, in addition to the calculations presented in the main article, we also take three Gaussian magma flow profiles above the locations of the major dehydration reactions of the slab, with a position, magnitude and width suggested by Ref.~\cite{Wilson2014}. We also consider the effect of all three sources combined.
\begin{figure*}
\includegraphics[width=14.44cm]{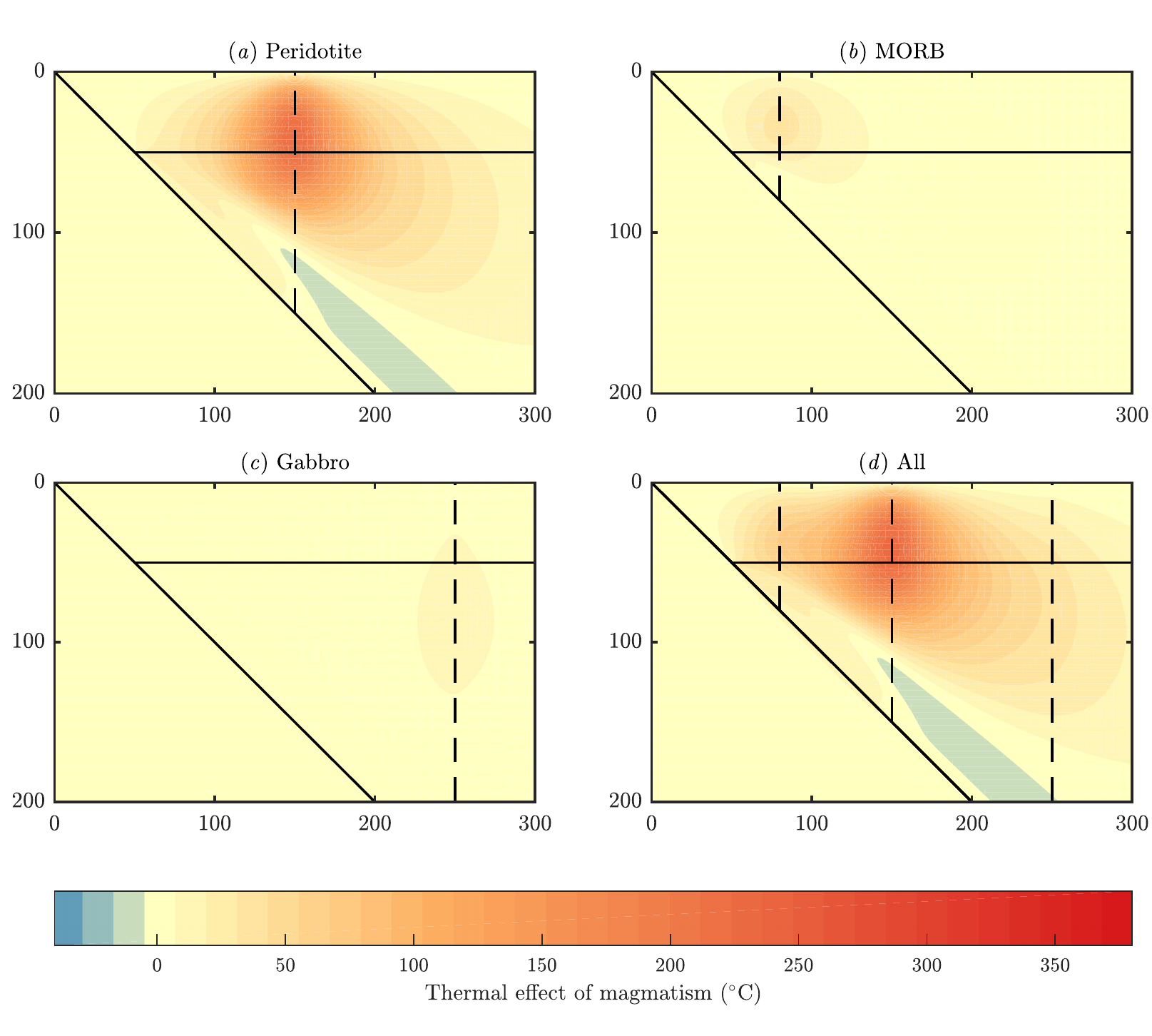}
\caption{The thermal impact of magmatism ($T-T_{\mathrm{ref.}}$) associated with the dehydration of (\textit{a}) peridotite, (\textit{b}) MORB,  and (\textit{c}) gabbros. We also show (\textit{d}) results when all three sources are combined. We indicate the individual sources as dashed lines at the centre of each Gaussian pulse. Horizontal and vertical scales are distance from the trench (units km). 
} \label{fig:wedge_sources}
\end{figure*}

As in the calculations in the main text, the principal result is that advective transport by magma substantially alters the thermal structure of subduction zones, as shown in Fig.~\ref{fig:wedge_sources}. Flow associated with the peridotite source (\textit{a}) is the most thermally significant, raising temperatures by over 200~K. Flow associated with the MORB source  (\textit{b}) raises temperatures near the trench by about 40~K; the gabbro source (\textit{c}) is thermally insignificant. The peridotite source is strongest because it is associated with the largest magma flux. The MORB source is weaker because the flux is smaller and also because it occurs nearer the trench than the peridotite source, which means that the mantle wedge above the MORB source is slightly cooler. The gabbro source is especially weak because the flux is smaller, and because it is narrower than the other sources, and so tends to diffuse laterally more strongly. The combined set of sources (\textit{d}) is dominated by the peridotite source, although there are also slightly elevated temperatures in the fore-arc region associated with the MORB source.

\clearpage

\end{document}